\documentclass[sigconf]{aamas} 

\usepackage{balance} 
\usepackage{algorithm}
\usepackage{algorithmic}
\usepackage{amsmath}
\usepackage{tabularx, booktabs, multirow, array}
\newcolumntype{Y}{>{\centering\arraybackslash}X}

\usepackage{pifont}
\newcommand{\cmark}{\ding{51}} 
\newcommand{\xmark}{\ding{55}} 

\usepackage{balance} 

\doi{IEON9331}

\makeatletter
\gdef\@copyrightpermission{
  \begin{minipage}{0.2\columnwidth}
   \href{https://creativecommons.org/licenses/by/4.0/}{\includegraphics[width=0.90\textwidth]{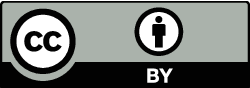}}
  \end{minipage}\hfill
  \begin{minipage}{0.8\columnwidth}
   \href{https://creativecommons.org/licenses/by/4.0/}{This work is licensed under a Creative Commons Attribution International 4.0 License.}
  \end{minipage}
  \vspace{5pt}
}
\makeatother

\setcopyright{ifaamas}
\acmConference[AAMAS '26]{Proc.\@ of the 25th International Conference
on Autonomous Agents and Multiagent Systems (AAMAS 2026)}{May 25 -- 29, 2026}
{Paphos, Cyprus}{C.~Amato, L.~Dennis, V.~Mascardi, J.~Thangarajah (eds.)}
\copyrightyear{2026}
\acmYear{2026}
\acmDOI{}
\acmPrice{}
\acmISBN{}

\title[AAMAS-2026 Formatting Instructions]{DEpiABS: Differentiable Epidemic Agent-Based Simulator}

\author{Zhijian Gao}
\affiliation{
  \institution{Nanyang Technological University}
  \city{Singapore}
  \country{Singapore}}
\email{zhijian001@e.ntu.edu.sg}

\author{Shuxin Li\textsuperscript{*}}
\affiliation{
  \institution{Nanyang Technological University}
  \city{Singapore}
  \country{Singapore}}
\email{shuxin.li@ntu.edu.sg}

\author{Bo An}
\affiliation{
  \institution{Nanyang Technological University}
  \city{Singapore}
  \country{Singapore}}
\email{boan@ntu.edu.sg}

\begin{abstract}
The COVID-19 pandemic highlighted the limitations of existing epidemic simulation tools. These tools provide information that guides non-pharmaceutical interventions (NPIs), yet many struggle to capture complex dynamics while remaining computationally practical and interpretable. We introduce DEpiABS, a scalable, differentiable agent-based model (DABM) that balances mechanistic detail, computational efficiency and interpretability. DEpiABS captures individual-level heterogeneity in health status, behaviour, and resource constraints, while also modelling epidemic processes like viral mutation and reinfection dynamics. The model is fully differentiable, enabling fast simulation and gradient-based parameter calibration. Building on this foundation, we introduce a z-score-based scaling method that maps small-scale simulations to any real-world population sizes with negligible loss in output granularity, reducing the computational burden when modelling large populations. We validate DEpiABS through sensitivity analysis and calibration to COVID-19 and flu data from ten regions of varying scales. Compared to the baseline, DEpiABS is more detailed, fully interpretable, and has reduced the average normal deviation in forecasting from 0.97 to 0.92 on COVID-19 mortality data and from 0.41 to 0.32 on influenza-like-illness data. Critically, these improvements are achieved without relying on auxiliary data, making DEpiABS a reliable, generalisable, and data-efficient framework for future epidemic response modelling.
\end{abstract}

\keywords{Epidemiology; Differentiable Agent-Based Model}

\newcommand{\BibTeX}{\rm B\kern-.05em{\sc i\kern-.025em b}\kern-.08em\TeX}

\begin{document}


\pagestyle{fancy}
\fancyhead{}


\maketitle 
\renewcommand{\thefootnote}{*}
\footnotetext{Corresponding author.}


\section{Introduction}

The COVID-19 pandemic exposed critical vulnerabilities in global public health systems, demonstrated by delayed and inappropriate implementation of non-pharmaceutical interventions (NPIs)~\cite{mohammadi2022human}. The effectiveness of interventions depends fundamentally on the quality of information available to policymakers, such as the expected infection and mortality trends, to which epidemic modelling methods contribute immensely. To provide \textit{insightful}, \textit{timely}, and \textit{trustworthy} information, epidemic models must represent real-world epidemic dynamics \textit{faithfully}~\cite{heesterbeek2015modeling}, \textit{expeditiously}~\cite{bershteyn2022real}, and \textit{transparently}~\cite{pokutnaya2023implementation}. Given the inadequacies in the control and mitigation of COVID-19 recorded in multiple countries~\cite{francisco2020understanding, khosrawipour2020failure, hallal2021overcoming}, the development of epidemic models with \textit{comprehensive advantages} over existing models in all 3 criteria, namely \textit{simulation fidelity}, \textit{computational efficiency}, and \textit{mechanistic interpretability}, is crucial for preparing against future epidemics. 

These criteria are inherently in tension, bringing trade-offs that have been addressed differently by multiple model paradigms and design techniques (\textbf{Table 1}), but never truly reconciled. \textit{Traditional models} such as SIR~\cite{kermack1927contribution} prioritise \textit{efficiency over fidelity}, and researchers have attempted to enrich them with features like age structure or network-based transmission to improve realism~\cite{salje2020estimating, gosce2020modelling}. \textit{Agent-based models (ABMs) reverse this priority} by explicitly representing individual heterogeneity; for example, Covasim~\cite{kerr2021covasim} simulates disease transmission through diverse social networks while optimising runtime and memory usage to retain feasible efficiency. More recent advances attempt \textit{introducing differentiability into ABMs}, enabling GPU acceleration and gradient-based optimisation without significant losses in realism. Two main approaches have emerged: \textit{ABM with neural surrogates}, and \textit{differentiable ABMs (DABMs)}. \textit{ABMs with neural surrogates} approximate model dynamics using artificial neural networks (ANN)~\cite{angione2022using, anirudh2022accurate}, boosting computational efficiency without losing simulation fidelity, yet \textit{sacrificing mechanistic interpretability}. \textit{DABM, as a paradigm}, reformulates model dynamics as tensor operations that preserve all details, therefore \textit{theoretically reconciles} the existing trade-offs. However, most existing DABMs~\cite{zhang2024ai, chopra2021deepabm, chopradifferentiable} integrate \textit{neural components} to infer epidemiological parameters from auxiliary data, which reintroduces \textit{opacity}. In conclusion, all existing models have non-negligible drawbacks in at least one criterion.

\begin{table*}[h]
    \centering
    \scriptsize
    \renewcommand{\arraystretch}{1.1}
    \setlength{\tabcolsep}{2pt}
    \begin{tabularx}{\linewidth}{
      l l
      *{4}{>{\centering\arraybackslash}X}  
      >{\centering\arraybackslash}X        
      >{\centering\arraybackslash}X  
    }
    \toprule
      \multirow{2}{*}{\textbf{Paradigm}} &
      \multirow{2}{*}{\textbf{Model}}     &
      \multicolumn{4}{c}{\textbf{Simulation Fidelity}} &
      \multicolumn{1}{c}{\textbf{Mechanistic Interpretability}} &
      \multicolumn{1}{c}{\textbf{Computational Efficiency}} \\
      \cmidrule(lr){3-6}\cmidrule(lr){7-7}\cmidrule(lr){8-8}
      & &
      Contact Heterogeneity & Behavioural Autonomy Simulation & Biological Process Modelling & Real Data Validation &
      Full Transparency &
      Runtime (1M population, 30 days) \\
    \midrule
    \multirow{2}{*}{\textbf{Non-ABM}}
      & FIGI-Net~\cite{song2025fine} & \xmark & \xmark & \xmark & \cmark  & \xmark & $<$1s \\
      & SIR~\cite{kermack1927contribution} & \xmark & \xmark & \xmark & \cmark & \cmark & $<$1s \\
    \midrule
    \multirow{3}{*}{\textbf{ABM}}
      & COVID-ABS~\cite{silva2020covid} & \cmark & \cmark & \xmark & \xmark & \cmark & $>$10 days \\
      & Covasim~\cite{kerr2021covasim} & \cmark & \xmark & \xmark & \cmark & \cmark & $<$1min \\
      & CityCOVID~\cite{ozik2021population} & \cmark & \cmark & \cmark & \cmark & \cmark & $<$20s (HPC) \\
      & ABM+NN~\cite{cozzi2025learning} & \cmark & \cmark & \xmark & \cmark & \xmark & $<$1s \\
    \midrule
    \multirow{3}{*}{\textbf{DABM}}.
      & DeepABM~\cite{chopra2021deepabm} & \cmark & \xmark & \xmark& \xmark & \xmark & $<$5min \\
      & JUNE-NZ~\cite{zhang2024ai} & \cmark & \xmark & \xmark & \cmark & \xmark & unspecified \\
      & GradABM~\cite{chopradifferentiable} & \cmark & \xmark & \xmark & \cmark & \xmark & $<$10s \\
      & DEpiABS (ours) & \cmark & \cmark & \cmark & \cmark  & \cmark & $<$1.5s  \\
    \bottomrule
    \end{tabularx}
    \caption{Comparison of modelling paradigms and representative models across key evaluation criteria. Biological processes modelling refers to the modelling of epidemic dynamics beyond transmission, including mutation and symptom progression.}
\end{table*}

Therefore, we propose \textit{DEpiABS}, a \textit{structure-centric} DABM that achieves \textit{comprehensive advantages} and fully leverages the theoretical potential of the paradigm. DEpiABS includes granular features such as \textit{resource-constrained behaviours, viral mutation, and reinfection} with key mechanisms validated via \textit{sensitivity analysis}, ensuring \textit{high simulation fidelity}. The \textit{associated computational cost} is reduced through \textit{differentiable approximation and tensorisation}. Relying solely on its ANN-free, \textit{fully transparent structure}, the model achieves \textit{global mechanistic interpretability}. We further optimise the design so that runtime scales \textit{linearly} with both population size and simulation length, and introduce a novel \textit{z-score-based scaling method} that maps outputs from fixed-size populations to any scales while preserving statistical patterns, effectively \textit{decoupling simulation cost from population size}. Finally, we validate DEpiABS’s forecasting ability across \textit{multi-regional} and \textit{multi-epidemiological} datasets. By outperforming \textit{GradABM} without using auxiliary data, we demonstrate superior \textit{forecasting ability} and \textit{data efficiency}. In summary, our contributions are:

\begin{itemize}
\item We introduce \textit{DEpiABS, a fine-grained and fully interpretable DABM designed around a novel structure-centric principle}, that is scalable to any population sizes and outperforms the state-of-the-art in forecasting on multi-regional, multi-disease data with higher data efficiency, demonstrating comprehensive advantages over existing models in all three criteria. 
\item We resolve the scalability challenge of ABMs with negligible compromise in realism by \textit{combining end-to-end differentiability with a z-score-based scaling method}, which enables simulating epidemics in large populations using a significantly smaller model population size.
\item We demonstrate that \textit{models under structure-centric design that prioritises feature granularity over data volume} can achieve performance comparable to, or even exceeding, data-centric benchmarks built around the opposite trade-off.
\end{itemize}


\section{Related Work}

\subsection{Traditional models}
Traditional models generally fall into three categories—compartmental, metapopulation, and statistical~\cite{perra2021non}. Compartmental models in the paradigm of SIR~\cite{kermack1927contribution} are popular for their tractability~\cite{alcaraz2012modeling, chen2015mathematical, cooper2020sir} but are deemed oversimplified~\cite{anderson1991infectious}. Metapopulation models introduce geographic structures~\cite{pei2020differential, costa2020metapopulation, coletti2021data} that enable regional analyses, but still overlook individual heterogeneity~\cite{ayabina2025note}. Statistical models emphasise predictive accuracy~\cite{flaxman2020estimating, bo2021effectiveness, song2025fine}, yet lack causal interpretability and do not explicitly model the epidemic processes. While all have provided valuable insights, they remain limited in realistically capturing complex epidemic dynamics.

\subsection{Agent-based models}
Agent-based models (ABMs) simulate disease spread through agents in structured environments~\cite{ajelli2010comparing} and are highlighted for their realism. They have supported epidemic planning since H1N1~\cite{lee2010simulating}. Recent ABMs have incorporated increasingly realistic components, such as economic structures~\cite{silva2020covid}, digital tracing~\cite{hinch2021openabm}, and psychological modules~\cite{mitsopoulos2023psychologically}. However, such realism incurs substantial computational cost, as individuals and their interactions must be modelled explicitly. Large-scale models like CityCOVID~\cite{ozik2021population} thus require supercomputing resources. One remedy is static scaling, which aggregates multiple individuals into representative agents~\cite{wise2022importance} but sacrifices output granularity by inflating rare events as scaling factors increase. Covasim~\cite{kerr2021covasim} alleviates this through dynamic scaling based on output magnitude using predefined thresholds, but this demands careful tuning to prevent discontinuities between adjacent data points. Computational burden is further amplified during calibration, with the result that formal calibration is frequently omitted in many models~\cite{silva2020covid}. Some models accelerate calibration with meta-heuristic algorithms~\cite{lestari2011pandemic, cockrell2018examining}, but yield limited improvement.

To improve efficiency, surrogates have been introduced. Compartmental and Bayesian emulators~\cite{fonseca2025optimal, vernon2022bayesian} accelerate simulation, but often oversimplify the dynamics of ABMs. ANN-based surrogates~\cite{angione2022using, anirudh2022accurate, madhavan2022deep, cozzi2025learning} better approximate dynamics and enable efficient gradient-based calibration, but at the cost of mechanistic interpretability.

\subsection{Differentiable agent-based models}
DABMs achieve end-to-end differentiability through smooth approximations and tensorisations, enhancing computational efficiency while preserving mechanistic interpretability, as it avoids opaque ANN modules. The minor fidelity loss from smooth approximations is typically negligible. However, most DABMs are data-centric models that employ inferential ANNs to utilise auxiliary data, reintroducing black-box structures. One of the first epidemiological applications, DeepABM~\cite{chopra2021deepabm}, models agent interactions with a graph neural network (GNN) and achieves a 200-fold acceleration over equivalent Mesa implementations, but lacks calibration. A subsequent model, JUNE-NZ~\cite{zhang2024ai}, integrates GNNs for contact structures with LSTM networks for temporal dynamics, fitting the 2019 New Zealand measles outbreak in Auckland well but showing limited generalisability. Besides, they do not model agent autonomy or biological processes. The most outstanding model so far is GradABM~\cite{chopradifferentiable} and its variants, which captures at least as many details as expert-designed ABMs~\cite{romero2021public, abueg2021modeling} and outperforms them on COVID-19 mortality and influenza-like-illness data across ten Massachusetts counties. 


\section{Problem Description}
This paper aims to contribute to \textit{pandemic mitigation} by addressing the inadequacies in information-providing epidemic models that support the design and implementation of intervention policies. Successes in mitigating the COVID-19 pandemic through intervention policies before biological and medical research released sufficient results had averted hundreds of thousands of cases in regions like South Korea~\cite{lee2020testing} and China~\cite{tian2020investigation}. Summarising from the literature, an adequate epidemic model should have:
\begin{itemize}
    \item \textbf{Simulation fidelity:} The model represents the real-world epidemic dynamics \textit{faithfully} to provide information useful for understanding and control~\cite{heesterbeek2015modeling}.
    \item \textbf{Computational efficiency:} The model executes \textit{expeditiously} to cope with the time constraints of a public health emergency~\cite{bershteyn2022real}.
    \item \textbf{Mechanistic interpretability:} The model operates \textit{transparently} to ensure the provided information is trustworthy to decision-makers~\cite{pokutnaya2023implementation}.
\end{itemize}

The challenge that remains open lies in designing a model \textit{without significant shortcomings} with respect to the three key criteria, because the properties that enhance one tend to degrade another: fidelity demands \textit{detail}, efficiency demands \textit{abstraction}, and interpretability \textit{restricts the use of black-box structures} that can dissolve the tension between the other two. 

\textit{Traditional models}~\cite{kermack1927contribution, coletti2021data} and \textit{simple ABMs}~\cite{kerr2021covasim, hinch2021openabm} are \textit{computationally efficient}, but their \textit{poor simulation fidelity} limits the reliability of the information they provide~\cite{thompson2022framework, teerawattananon2022recalibrating}; \textit{complex ABMs}~\cite{silva2020covid, ozik2021population} give \textit{faithful simulations}, but their \textit{poor computational efficiency} can make timely simulation and calibration costly, and sometimes impossible~\cite{hunter2022validating, berec2023contact}; \textit{ABM with NN surrogates}~\cite{anirudh2022accurate, madhavan2022deep} are good in both criteria, but they \textit{compromise mechanistic interpretability}. The \textit{DABM} paradigm holds significant promise for addressing this challenge, as it does not possess non-negligible shortcomings on a paradigmatic level. However, existing \textit{data-centric DABMs}~\cite{chopra2021deepabm, zhang2024ai, chopradifferentiable} remain limited in \textit{simulation fidelity} and \textit{mechanistic interpretability} due to \textit{specific design choices}. 

Therefore, we aim to fully leverage the paradigm's potential by following a novel \textit{structure-centric DABM-design principle}. We first construct a \textit{fine-grained agent-based model} that incorporates extensive details without regard for computational cost, thus ensuring \textit{simulation fidelity}. We then convert it into a \textit{DABM} through differentiable approximations to accelerate computation and parameter optimisation. Finally, we propose a \textit{z-score-based scaling method} that enables simulation under large data–model scale differences (up to 2000 times) while preserving output granularity, thus ensuring \textit{computational efficiency}. Unlike data-centric DABMs, we \textit{reject black-box components}, thus ensuring \textit{mechanistic interpretability}.


\section{Model Structure}
In this section, we demonstrate how we operationalise \textit{simulation fidelity} and \textit{mechanistic interpretability} as the core of our novel structure-centric design. The model is constructed from fine-grained, fully transparent mechanisms that explicitly mirror real-world epidemiological and social processes. It contains three components: \textbf{Society}, which includes all non-epidemiological context configurations; \textbf{Epidemic}, which includes all epidemiological configurations; and \textbf{Population}, the configurations of the agents. 

\subsection{Society}
The Society component defines the model’s foundational operational context through three modules: the \textbf{temporal horizon}, \textbf{spatial context}, and \textbf{a simple economic system}.

The temporal horizon is discretised into daily time steps, matching the granularity of most epidemiological surveillance data. Each day is labelled as a weekday or weekend and grouped into months using modulo operations. For simplicity, each month is approximated as 30 days, which avoids unnecessary structural complexity while preserving temporal periodicity.

The spatial context, following existing ABMs, is a map composed of three facility types: offices~\cite{hinch2021openabm, kerr2021covasim}, markets~\cite{sarbajna2024covid19, nitzsche2024agent}, and hospitals~\cite{pham2021interventions, hinch2021openabm}. These facilities capture the main venues of employment, consumption, and healthcare, supporting three interaction types: within-office, worker–customer, and doctor–patient. The map contains multiple clusters, each comprising one of each facility type, and the number of clusters is configurable to simulate polycentric urban planning~\cite{derudder2022polycentric}. Agents must select a location to visit, and encounters occur probabilistically among co-located agents. Staying at home is modelled as remaining outside the map and achieving full isolation~\cite{williams2023epidemic}.

The economic system builds on the above mechanisms. Agents balance income generation and expenditure: each receives a monthly salary and must pay recurring bills at the end of each month. Absenteeism accumulates when agents skip work on weekdays, and exceeding an employer-defined threshold results in a temporary salary reduction. Each day, agents consume one unit of essential supplies and can only purchase replacements in markets. These mechanisms jointly create financial incentives that may conflict with health-preserving behaviours, compelling agents to trade off infection avoidance against income maintenance, reproducing empirically observed presenteeism due to financial pressures~\cite{gallagher2021low, garcia2023presenteeism}.

\subsection{Epidemic}

\begin{figure}[t]
    \centering
\includegraphics[width=0.95\linewidth]{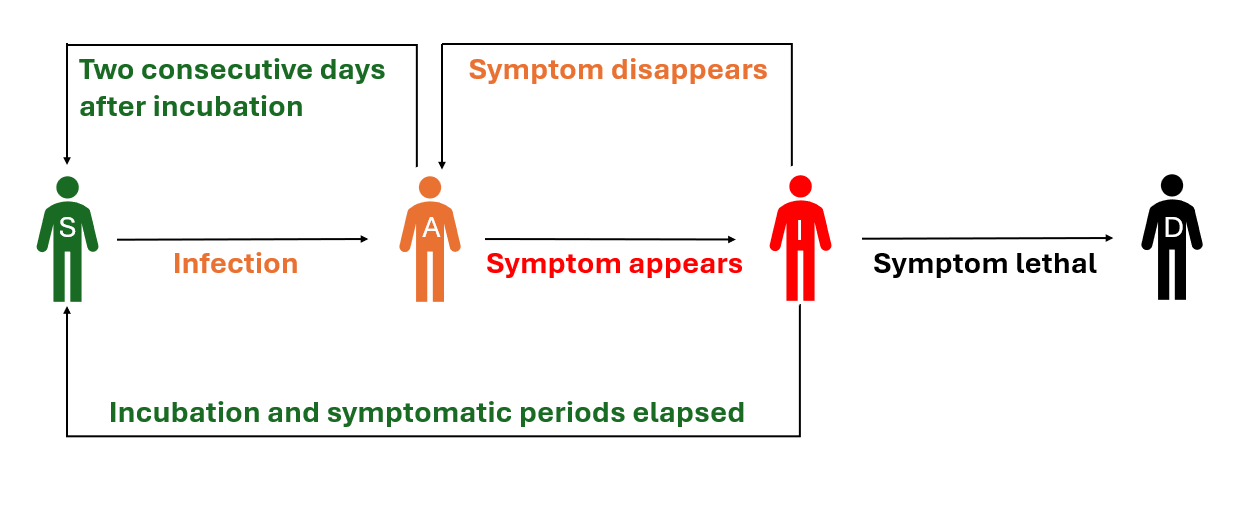}
    \caption{Health class transition}
    \label{healthclass}
\end{figure}

The Epidemic component captures disease dynamics across three levels: \textbf{between-host transmission, within-host progression}, and \textbf{pathogen mutation.} To represent these processes, we transform the classical $SEIR$ framework into an $SAID$ formulation that accounts for partial immunity and reinfection. Specifically, agents who recover from infection re-enter the $S$ (Susceptible) class. An additional Asymptomatic Infected ($A$) class is introduced to represent infectious carriers without symptoms, supported by empirical evidence of widespread asymptomatic transmission~\cite{kucirka2020variation, oran2021proportion}. The Exposed ($E$) class is omitted as exposure can be inferred directly from expected encounters among agents. The Infected ($I$) class is redefined to represent symptomatic individuals. A Deceased ($D$) class captures agents whose symptom severity surpasses a lethal threshold. The traditional Recovered ($R$) class is removed as permanent immunity is replaced by cumulative partial immunity. The transition is shown in \textbf{Figure \ref{healthclass}}.

The \textit{between-host transmission} mechanism governs how infections propagate through encounters between infectious and susceptible agents. Transmission occurs probabilistically upon contact, consistent with the biological process described by the dose–response model~\cite{haas2014quantitative, brouwer2017dose}. Building on subsequent analysis of infection initiation dynamics~\cite{teunis2000beta, mayer2011dynamic}, we model each encounter as a single \textit{infection-initiation} event, where infection occurs when cumulative infection initiations surpass the host’s immune tolerance. Accordingly, partial immunity is modelled by increasing the number of transmissions required for infection establishment.

The \textit{within-host progression} mechanism quantifies the temporal dynamics of symptom development for each infected agent. Following infection, an agent undergoes an incubation period (0 severity), followed by a symptomatic period. Agents naturally recover when the total incubation and symptomatic durations elapse. The incubation and symptomatic durations, denoted $t_{inc}^{j}$ and $t_{sym}^{j}$ for agent $j$ at time step $t$, are independently resampled for each infection episode based on empirical results reported in ~\cite{McAloone039652, PPR:PPR133359}:
\begin{equation}
    \begin{split}
        t_{inc}^{j} \sim \log\mathcal{N}(1.63, 0.5), 
        t_{sym}^{j} \sim \frac{age^{j}}{43.11 \cdot immune_{t-1}^{j}} \cdot \mathcal{N}(7.86, 6.46). \nonumber
    \end{split}
\end{equation}
Agents remain in class $A$ during incubation and move to $I$ once symptoms appear. Symptom severity follows a Normal distribution, a standard assumption for ordinal health outcomes~\cite{hopper1988random, cauchi2024individualized}, with age-dependent mean and variance reflecting greater severity and variability in older individuals~\cite{farshbafnadi2021aging}. If severity falls below zero during the symptomatic phase, the agent reverts to $A$, representing asymptomatic remission~\cite{kwong2016editorial}. Recovery occurs early after two consecutive non-positive days~\cite{kim2021duration}. Agents whose symptom severity exceeds 6 are classified as $D$. 

The \textit{pathogen mutation} mechanism introduces temporal variation in transmission and progression. At each time step, the pathogen mutates with a fixed probability. When a mutation occurs, agents with partial immunity lose one level of protection, allowing reinfection consistent with observed immune escape~\cite{nonaka2021genomic, chen2024does}, and symptom parameters $(\mu, \sigma)$ are rescaled to 0.5–1.5 times their prior values to capture variant-induced heterogeneity in symptom trajectories.

\subsection{Population}
The Population characterises agents along three aspects: \textit{basic configurations, observation interfaces,} and \textit{behavioural rules}. Shared structural forms define the population collectively, while parameter randomisation within these forms yields individual heterogeneity. 

The \textit{basic configurations} define each agent’s static attributes, including demographic and personal characteristics. Demographics cover age, workplace address, preferred market and hospital, income (with salary-reduction thresholds), and expenditure levels (modelled as fixed monthly bills). Personal characteristics include irrationality factors represented by randomised decision rules~\cite{hauber2016statistical} (\textbf{A.4}) and judgment thresholds~\cite{granovetter1978threshold} for interpreting observed situations, following common practices.

The \textit{observation interfaces} include an internal and an external interface. The internal observation interface retrieves the current savings and number of days of absenteeism, supplies (in daily quotas), and symptom severity, which inform judgment on financial, supply, and health states, respectively. The external observation interface observes the number of agents with each health class and aggregates them into the proportion of asymptomatic infections in the infected population, the proportion of cumulative infections in the whole population, and the proportion of cumulative infections resulting in death, which constitutes judgments on epidemic severity.

The \textit{behavioural rule} maps observations to location choices. Following the fast-and-frugal heuristics framework~\cite{gigerenzer2011heuristic, gigerenzer2004fast}, it consists of if–then rules that compare observations against judgment thresholds (basic configurations) and output a distribution over four decisions: \textit{stay at home, go to work, go shopping}, or \textit{go to the hospital}. Observations are first translated into drives (intensity levels 1, 2, or 3) to each decision, with the following decision-drive correspondences:
\begin{itemize}
    \item \textbf{Stay at home:} Epidemic severity and health state (mild).
    \item \textbf{Go to work:} Financial state.
    \item \textbf{Go shopping:} Supply state.
    \item \textbf{Go to the hospital:} Health state (severe).
\end{itemize}
The emergency level is determined by the extent to which the drive to one decision exceeds the others in intensity, while the decisions sorted with respect to their drives define a plausibility ranking that forms a priority heuristic, a widely used behavioural model\cite{birnbaum2008evaluation,brandstatter2006priority}.
Ties are resolved by prioritising shopping, hospital visits, staying home, and working, in that order, reflecting survival priorities across supplies, health, infection avoidance, and income, aligning with observed behaviours such as panic buying and delayed care-seeking during epidemics~\cite{CHEN202057, motamed2021factors}.

Subsequently, a distribution over the four decisions is determined using the initialised randomisation rules representing irrationality, which define three sets of probability mass functions with different skewness. Higher emergency levels correspond to more skewed distributions, simulating bounded rationality and decision salience~\cite{yeung2012metacognition}, and decisions with higher plausibility ranks will be assigned with larger probability masses. The decision will be sampled from the determined categorical distribution, and a location will be selected correspondingly. 

\subsection{Model Workflow}
\begin{figure}
    \centering
\includegraphics[width=\linewidth]{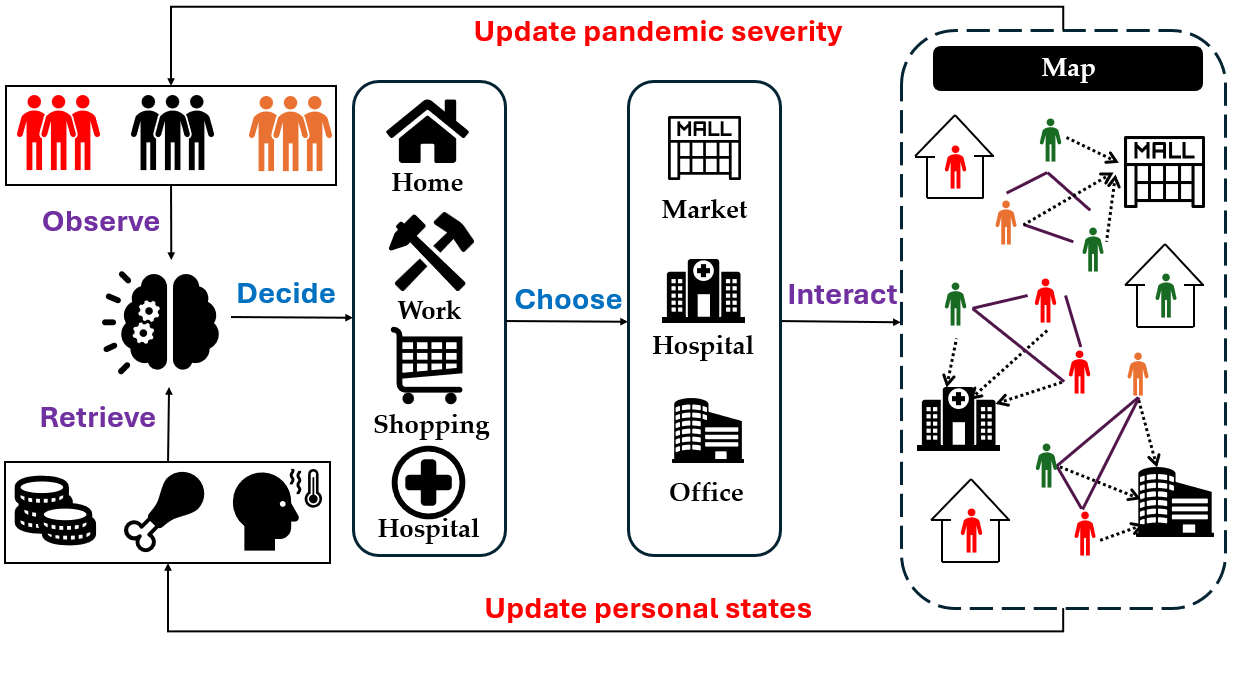}
    \caption{Model Workflow}
    \label{workflow}
\end{figure}

Integrating the components described above, the complete ABM workflow is illustrated in \textbf{Figure \ref{workflow}}. At each time step, every living agent observes epidemic severity via the external observation interface and retrieves its financial, supply, and health states via the internal interface. These inputs are processed by the behavioural rules to select a location on the map(or stay at home). Agents who go out may encounter others making the same choice. When a susceptible agent meets an infectious one, transmission occurs probabilistically, and infection is established once cumulative transmissions exceed the agent’s immune tolerance. Transmission outcomes and decisions will then jointly update each agent’s personal states. Then the ABM iterates this process over successive time steps until the simulation reaches the predefined temporal horizon.


\section{Parameter Calibration Method}
In this section, we present our parameter calibration method that aims to align the fine-grained ABM with real-world epidemic data while maintaining computational efficiency. Existing calibration methods for agent-based epidemic models often simplify the model structure, aggregate agent behaviours, or employ sampling approximations to reduce computational cost. Although effective for scaling, these approaches weaken the model’s mechanistic fidelity and make parameter calibration against empirical data unreliable. To address these limitations, we propose to \textit{combine end-to-end differentiability with a z-score-based scaling method}, forming a unified optimisation framework that preserves micro-level interpretability while enabling efficient data-driven calibration. Specifically, the differentiable approximation reformulates the ABM into a continuous computational graph that supports gradient-based parameter learning directly from empirical epidemic data. At the same time, the z-score-based scaling normalises and rescales outputs across populations to stabilise optimisation and generalise learned dynamics to large-scale epidemics. Together, these techniques establish a hybrid framework that seamlessly integrates fine-grained mechanistic simulation with data-driven calibration, ensuring that the ABM remains both computationally tractable and empirically faithful.

\subsection{Differentiable Approximation}
Most mechanisms in the ABM are inherently non-differentiable, such as the discrete decision rules and categorical state transitions that define agent behaviours. While some components, such as if–then decision rules, can only be relaxed, others, such as categorical variables (e.g., health classes), can be tensorised into differentiable representations. To construct \textit{an end-to-end differentiable structure that supports gradient-based calibration}, we approximate these discrete processes with differentiable surrogate functions. This reformulation transforms the original rule-based ABM into a differentiable computational graph, enabling GPU-accelerated parallelisation and allowing parameters to be calibrated directly against empirical epidemic data.


\subsubsection{Relaxation}
Three types of non-differentiable mechanisms require relaxation: \textbf{if–then rules, modulo functions, and random variables.} To this end, we introduce a novel \textit{selective relaxation method} to relax if-then rules, design a \textit{differentiable periodic indicator following the idea of sinusoidal mappings of time}~\cite{kazemi2019time2vec} to relax the modulo function, and use the \textit{reparametrisation trick}~\cite{kingma2013auto, jang2016categorical} to relax random variables.

For if-then rules (of the generic form “if $a>A$ then $x$,” equivalent to the Heaviside step function $x\mathbf{I}(a>A)$), relaxation inevitably trades off condition precision against optimisation stability, as closer approximations to the step function yield higher precision but increase the risk of vanishing gradients during optimisation. To balance this trade-off, we exploit the fact that different mechanisms demand different levels of precision. Our selective relaxation method applies distinct continuous relaxations under three categories: \textit{precision-critical, moderately precise, and fuzzy conditions.}
For precision-critical cases (e.g., health state transitions), we use:
\begin{equation}
    h(a)=x\cdot\frac{\text{ReLU}(a-A)}{\text{ReLU}(a-A) + \xi}\begin{cases}
        =0\text{ if }a\leq A\\
        \approx x\text{ otherwise} \nonumber
    \end{cases}
\end{equation}
where $\xi$ is a small slack variable. This closely approximates the Heaviside step function, but yields an extremely small gradient $x\frac{\xi}{(\mathbf{I}(a>A)+\xi)^2}\mathbf{I}(a>A)$. Therefore, for moderate-precision scenarios (e.g., symptom progression), we use a smoother variant:
\begin{equation}
    h(a)=x\cdot\text{ReLU}(\tanh(a-A))=\begin{cases}
        0\qquad\qquad\qquad\text{ if }a\leq A\\
        x\cdot\tanh(a-A)\text{ otherwise} \nonumber
    \end{cases}
\end{equation}
which still outputs zero when the condition is unmet, but increases smoothly when satisfied. Its gradient, $x\mathbf{I}(a>A)\text{sech}^2(a-A)\in[-|x|, |x|]$], help maintain better stability during gradient descent. For fuzzy scenarios such as behavioural decisions, we adopt the logistic approximation, following the literature~\cite{andelfinger2021differentiable}:
\begin{equation}
    h(a)=x\cdot\sigma(k \cdot (a-A)) \nonumber
\end{equation}
where $k$ is the steepness factor. It provides a continuous transition between 0 and 1, reflecting boundedly rational decision processes.

Modulo functions are used to make calendar-like time step classification into months and weekdays or weekends. Considering only integer $t$, we design the following differentiable periodic indicator as a relaxation:
\begin{equation}
    \rho(t;m,n)=\mathbf{I}(\cos^2(\frac{(t - m)\pi}{n})>\cos(\frac{\pi}{n}))=\begin{cases}
        1\text{ if }t \equiv m\pmod{n}\\
        0\text{ otherwise}
    \end{cases} \nonumber
\end{equation}
where the Heaviside step function is approximated using the precision-oriented relaxation from \textbf{(2)}. Using this function, weekends can be labelled by $\rho(t; 6, 7) + \rho(t; 0, 7)$ and month ends by $\rho(t; 0, 30)$.

Relaxation of random variables using the reparameterisation trick has been well studied and widely used in the field of Variational Auto-Encoders (VAEs)~\cite{kingma2013auto, rezende2014stochastic, jang2016categorical}. Since well-established reparameterisation techniques already exist for the random variables used in our model, i.e. the Normal and Categorical random variables (with the Bernoulli random variable treated as a special case of the latter), we directly adopt these standard formulations.

\subsubsection{Tensorisation}
The Tensorisation mainly follows established DABM practices (details in \textbf{A.2–A.4}). We elaborate on specific adjustments for three mechanisms: categorical variables, agents' encounters, and the transmission mechanism. Categorical variables are represented by one-hot or multi-hot tensors. Agents' encounters are tensorised as an adjacency representation, following the same idea as ~\cite{chopradifferentiable} but operationalised differently due to the difference in the model's spatial context. Transmission is then computed using the tensorised health classes and the encounter matrix.

Four types of categorical variables are tensorised: health classes, decisions, location choices, and location preferences. Health classes and decisions are represented by one-hot vectors $h_t^{j}\in\{0, 1\}^4$ and $d_t^{j}\in\{0, 1\}^3$ (staying at home excludes further interactions), respectively, and location choices are represented by one-hot matrices $l_t^{j}\in\{0, 1\}^{n\times3}$ with $n$ being the number of clusters of locations. Similarly, location preferences are represented by multi-hot tensors $L^{j}\in\{0, 1\}^{n\times3\times3}$ with three entries of 1, because there must be a preference (and for work, workplace) for each of the three decisions, excluding staying at home. 

Encounters are derived from location choice matrices $l_t^{j}$. Let $P$ be the population size and $m$ the encounter probability. The encounter matrix $E_t$ is defined as:
\begin{align}
        &l_{t_1}=\overset{P}{\underset{j=1}{\Sigma}}l_t^{j}, \quad
        l_{t_2}=\frac{1}{\sqrt{2}}(||l_t^{i}-l_t^{j}||_\mathcal{F})_{i=1, j=1}^{P, P}
        \nonumber \\ 
        &E_t=(l_{t_1}l_{t_1}^T\odot(\mathbf{1}_P-l_{t_2}))\odot(\mathbf{1}_P-I_P)\cdot m \nonumber
\end{align}
where $l_{t_1}$ indicates agents who leave home, and $l_{t_2}$ specifies whether agent pairs choose the same or different locations, or if one remains at home. Entry-wise multiplication of $l_{t_1}l_{t_1}^T$ and $(\mathbf{1}-l_{t_2})$ identifies co-located pairs, while $(\mathbf{1}_P-I_P)$ removes self-encounters. Multiplying by $m$ yields pairwise encounter probabilities.

Transmissions are then computed. Let $\beta$ denote the transmission probability and $h_t=(h_t^{j})_{j=1}^P\in\{0,1\}^{4\times P}$ the stacked health labels. The expected number of exposure events per healthy agent is:
\begin{equation}
    E_t^I=({E_t\odot h_{t-1}}_1)\cdot({h_{t-1}}_2+{h_{t-1}}_3)\cdot\beta \nonumber
\end{equation}
Here, ${E_t\odot h_{t-1}}_1$ selects encounters involving susceptible agents, while ${h_{t-1}}_2+{h_{t-1}}_3$ represents all asymptomatic and symptomatic infected agents. Their product gives expected close contacts, and multiplying by $\beta$ yields expected exposure events.

\subsection{Output Scaling}

While differentiable approximation significantly accelerates computations by enabling gradient-based parallelisation, it contributes little to reducing memory consumption. To further enhance computational efficiency, particularly with respect to memory, an additional mechanism is required. Output scaling addresses this by allowing a small simulated population to represent a much larger real-world one. However, existing methods are limited: static scaling, which scales between model and real populations, is too rigid and reduces output granularity when population differences are large, while dynamic scaling, which scales outputs solely by their magnitudes, demands careful hyperparameter tuning.

To overcome these limitations, we propose a novel z-score-based scaling method, a forecasting-oriented scaling method that directly scales between the model outputs and the empirical data, rather than between populations or with respect to output magnitudes. Forecasting-orientation is justified because forecasting, which relies on calibrating model outputs to real data, is the task most sensitive to scalability. Compared with static scaling, our method allows a larger difference between model and real population scales, since key epidemic indices (e.g., daily infections and mortalities) constitute only small fractions of the total population, which means that the ratio between a small-scale model output and empirical data is much smaller than the ratio between their populations. Compared with dynamic scaling, our approach requires no hyperparameter tuning, as demonstrated below.

Consider the model as a function $f(w; x)$, where $w$ represents the non-learnable parameters and $x$ represents the learnable parameters. Let the output time series be $\tilde{y}=f(w;x)$. Our scaling method will first compute the z-score 
$\tilde{y}_z=\frac{\tilde{y}-\tilde{y}_\mu}{\tilde{y}_\sigma}$
where $\tilde{y}_\mu$ and $\tilde{y}-\sigma$ are the mean and standard deviation of $\tilde{y}$. After that, $\tilde{y}_z$ will be rescaled through an inverse computation of z-score via $y_\mu$ and $y_\sigma$, the mean and standard deviation of the real data for calibration, and translated so that the smallest entries of the transformed output and the original output can align, i.e.
\begin{equation}
        \tilde{y}'=\tilde{y}_zy_\sigma+y_\mu, \quad
        \hat{y}=\tilde{y}'-\tilde{y}'_{min}\mathbf{1}+y_{min}\mathbf{1} \nonumber
\end{equation}
The translation ensures non-negative values and aligns the minimum of the transformed output with that of the real data. 

This method directly matches the first two statistical moments (scale and location) of the simulated and real series, guaranteeing a consistent and reliable projection of the model output without distorting its trend, yet the transformation parameters are automatically derived from statistical properties, thus eliminating the need for manual hyperparameter tuning. 


\section{Experiment}
In this section, we demonstrate approaches we take to further justify our model's simulation fidelity and computational efficiency. For simulation fidelity, we conducted sensitivity analysis to validate the realism of the model's key mechanisms, and performance evaluation using GradABM~\cite{chopradifferentiable} as a baseline to demonstrate the model's forecasting capability, generalisability, and data efficiency. For computational efficiency, we demonstrate how the model's runtime varies with population size and forecast horizon.

\subsection{Sensitivity Analysis}
\subsubsection{One-at-a-time Analysis}
This analysis shows how key simulations vary when parameters involved in the related mechanisms are changed. The evaluated simulations include the cumulative infections, cumulative deaths, and the number of infected agents in critical conditions (defined as agents with symptom severities over 4). For cumulative infections, we examine the influences of transmission probability, mean monthly bill and maximum amount of daily purchasable supplies, which are key factors of the pathogen's infectiousness, financial pressure, and supply stability, respectively. For the cumulative deceased and the number of agents in critical condition, we examined the influences of the mean and standard deviation of daily severity changes. As illustrated in \textbf{Figures 3, 4, 5}, key mechanisms of the model operate realistically: the infection growth rates increase as the pathogen's infectiousness and financial pressure increase, and as the supply stability decreases.
\begin{figure}
    \centering
    \includegraphics[width=\linewidth]{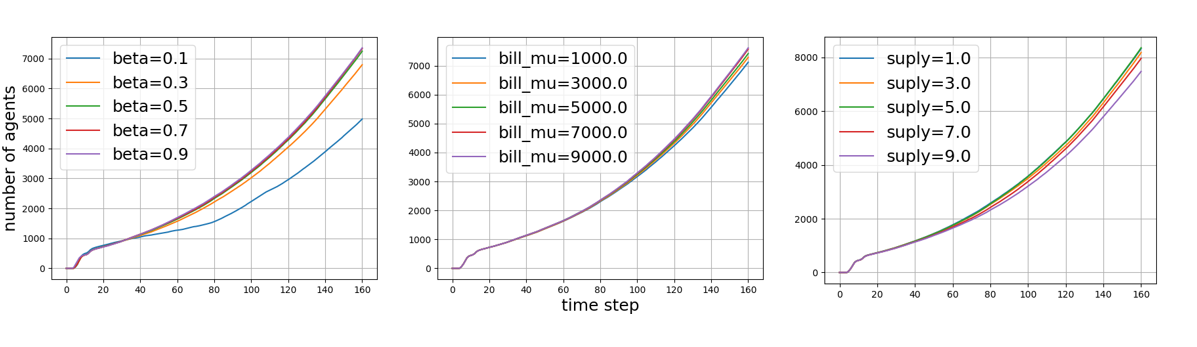}
    \caption{Cumulative infections under different transmission probabilities (left), mean monthly bill (middle), and maximum purchasable daily quotas of supplies (right)}
\end{figure}
\begin{figure}
    \centering
    \includegraphics[width=\linewidth]{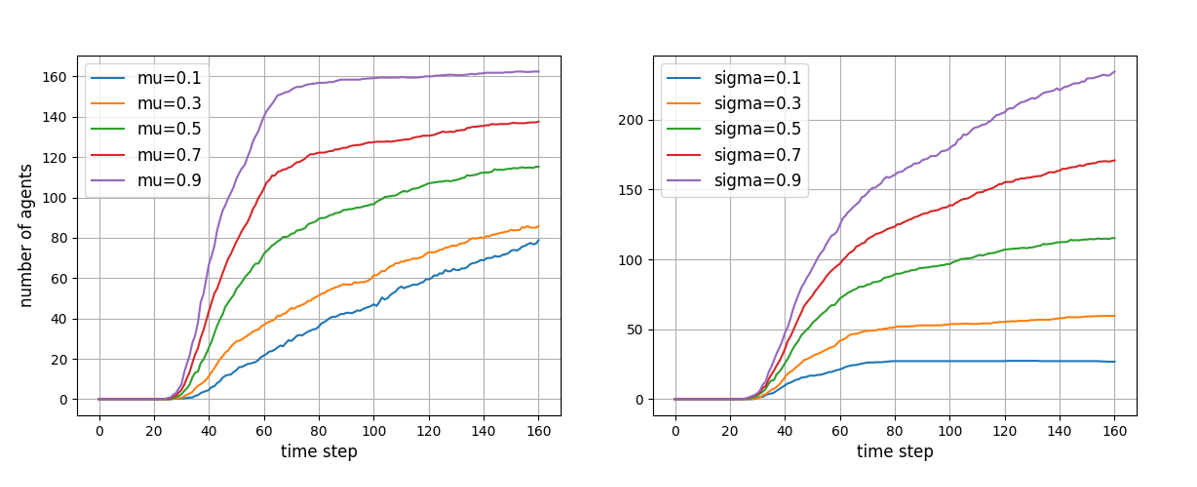}
    \caption{Cumulative mortality under different means (left) and standard deviations (right) of symptom progression}
\end{figure}
\begin{figure}
    \centering
    \includegraphics[width=\linewidth]{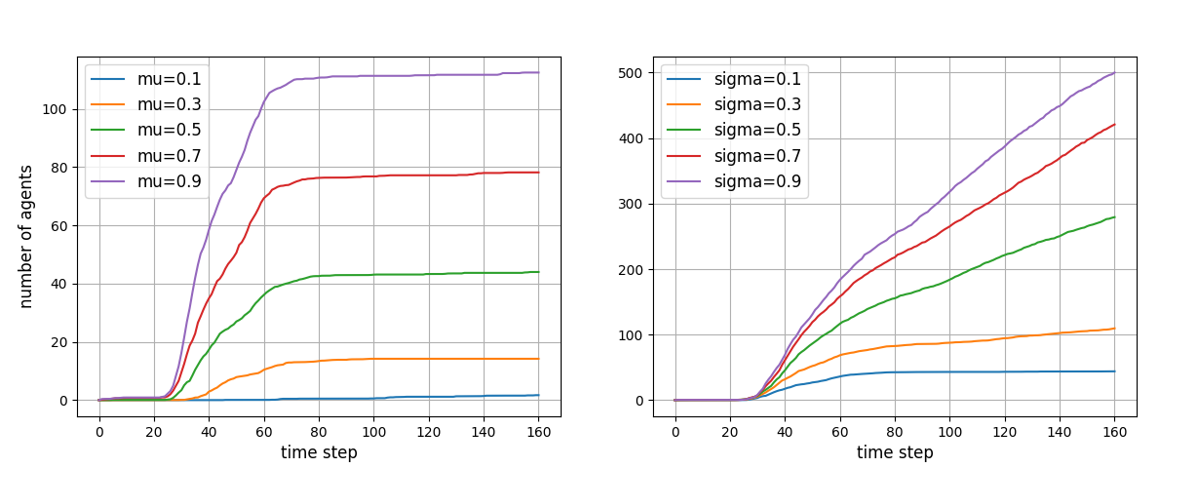}
    \caption{Cumulative agents in critical conditions under different means (left) and standard deviations (right) of symptom progression}
\end{figure}

\subsubsection{Sobol Analysis}
To explore the influences of the mutual effects of the parameters, we employed the Sobol variance decomposition method~\cite{saltelli2000sensitivity}, using the total-effect Sobol indices with 95\% CI, and Saltelli sampling size $N(2d+2)$, $N=1024$.

The sensitivity analysis demonstrated a clear and expected hierarchy in parameter influences on epidemic dynamics, even with mutual effects considered. For simulated infections, transmission and encounter probabilities dominate with Sobol indices at least 45\% higher than other groups, consistent with their fundamental role of impact in disease spread. Health progression parameters and behavioural parameters form a secondary tier, reflecting the model's dual biological-behavioural architecture. For simulated deceased cases, parameters associated with individual health emerge as the dominant factors, demonstrating that the within-host progression mechanisms function as intended, given that agent mortality depends solely on these health dynamics. 

We also compared the influence of the parameters on the number of agents making each decision. The decision-making parameters prove most influential across all outputs except for the number of working agents. In this latter case, transmission parameters retain dominance, albeit narrowly surpassing decision-making factors. The persistent impact of transmission and health-related parameters aligns with expectations for epidemic-driven decision-making. An unexpected finding concerns the differential impact of health-related parameters on decision-making: their effect on shopping and staying home appears stronger than on hospital visits. A preliminary hypothesis attributes this to overly restrictive parameter ranges to determine mean symptom severity change, resulting in too few agents developing symptoms severe enough to warrant hospitalisation. To investigate this further, the range was expanded in subsequent experiments. After allowing parameters that represent the spread of more fatal diseases to be sampled, the influences of health-related parameters become reasonable: the health-related factors are most influential to the number of agents going to the hospital, and second most influential to the number of agents staying at home.

\subsection{Performance Evaluation}
\subsubsection{Baseline}
We use the three versions of GradABM~\cite{chopradifferentiable} as baselines. GradABM is a differentiable agent-based model designed to efficiently simulate large-scale epidemic dynamics and is among the most widely cited open-source DABMs with well-documented calibration and forecasting results. It was validated using COVID-19 mortality data from 10 counties in Massachusetts, with population scales from 70000 to 1600000, and influenza-like illness (ILI) data for the state. The three versions are: C-GradABM, which doesn't use any auxiliary data; DC-GradABM, which uses a deep neural network to infer simulation parameters from regional heterogeneous features such as mobility and census data; and JDC-GradABM, which generalises this approach by leveraging heterogeneous data from all regions jointly. They all use 7 parameters for the COVID-19 simulation and 8 parameters for the influenza simulation. In order to make a reliable comparison, we adopted the same data, data processing method, evaluation lengths, and metrics. More details are explained in \textbf{B.2}.

\subsubsection{Results}
\begin{table*}[h]
    \centering
    \begin{tabular}{c|ccc|ccc|}
        \toprule
         & \multicolumn{3}{c|}{\textbf{COVID-19}} & \multicolumn{3}{c|}{\textbf{Flu}} \\
        \cmidrule(lr){2-4} \cmidrule(lr){5-7} 
        \textbf{Model} & \textbf{ND} & \textbf{RMSE} & \textbf{MAE} & \textbf{ND} & \textbf{RMSE} & \textbf{MAE} \\
        \midrule
        C-GradABM & $2.39\pm0.35$  & $205.24\pm42.56$ & $73.66\pm10.88$ & $0.88\pm0.14$ & $2.97\pm0.44$ & $2.64\pm0.43$ \\
        \hline
        DC-GradABM & $1.15\pm0.24$  & $67.09\pm23.89$ & $35.50\pm7.37$ & $0.50\pm0.19$ & $1.78\pm0.62$ & $1.50\pm0.57$ \\
        \hline
        JDC-GradABM & $0.97\pm0.18$  & $50.99\pm12.12$ & $30.02\pm5.60$ & $0.41\pm0.02$ & $1.47\pm0.06$ & $1.22\pm0.06$ \\
        \hline
        \textbf{DEpiABS} & $\textbf{0.92}\pm\textbf{0.05}$  & $\textbf{18.95}\pm\textbf{4.39}$ & $\textbf{12.97}\pm\textbf{2.25}$ & $\textbf{0.32}\pm\textbf{0.05}$ & $\textbf{1.80}\pm\textbf{0.28}$ & $\textbf{1.47}\pm\textbf{0.23}$ \\
        \bottomrule
    \end{tabular}
    \caption{Comparison between the performances of GradABM and DEpiABS. Given real and simulated time series $y$ and $\hat{y}$ of length $T$, the metrics used are defined as follows: \textbf{ND}$=\frac{\Sigma_{t=1}^T|y_t-\hat{y}_t|}{\Sigma_{t=1}^T|y_t|}$, \textbf{RMSE}$=\frac{1}{T}\sqrt{\Sigma_{t=1}^T|y_t^2-\hat{y}_t^2|}$, \textbf{MAE}$=\frac{1}{T}\Sigma_{t=1}^T|y_t-\hat{y}_t|$}
\end{table*}
\textbf{Table 2} demonstrates the results obtained by our model and the comparisons to the results of the baselines. For COVID-19 mortality, the mean and standard deviation of forecasting errors over 5 runs given by DEpiABS are both smaller than JDC-GradABM, the most powerful version of GradABM, in all metrics. For influenza, DEpiABS gives a smaller mean normal deviation than JDC-GradABM, and mean RMSE and MAE similar to DC-GradABM. The standard deviations of all norms are between JDC-GradABM and DC-GradABM. These results demonstrate our model's good forecasting ability and generalisability. We also illustrate the comparisons between the model's forecast and the real data of Norfolk County in \textbf{Figure 6}. Illustrations of results on other regions can be found in \textbf{B.2}. 
\begin{figure}
    \centering
    \includegraphics[width=\linewidth]{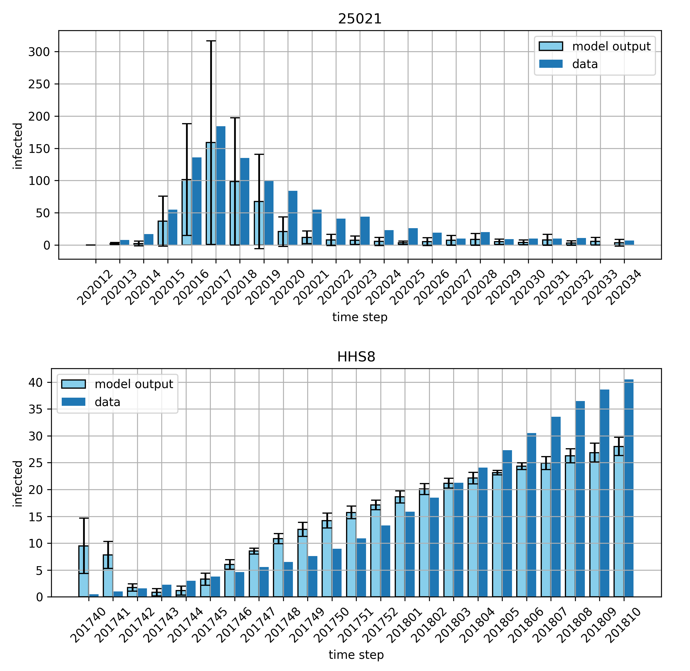}
    \caption{Forecasted and real COVID mortality (top) and forecasted and real ILI (bottom) in Norfolk County, Massachusetts}
\end{figure}

This yields three key insights. First, our model is data-efficient, as it achieves performance comparable to or exceeding JDC-GradABM, the strongest variant using aggregated auxiliary data from all counties, while using only the same amount of data as C-GradABM, the weakest variant without auxiliary data. Second, our z-score-based scaling method proves reliable, as the reported results are obtained from a 500-agent simulation employing this method. Third, our structure-centric DABM-design principle is competitive, since our model, one of the first of its kind, outperforms the state-of-the-art data-centric DABM.

\subsection{Scalability Test}
With our z-score-based method, the population size required to maintain reasonable output granularity is low, but a larger population size is still beneficial for making more accurate forecasts. Thus, the scalability test is important. Following the evaluation method incorporated by many existing works~\cite{chopra2021deepabm, chopradifferentiable, kerr2021covasim}, we conducted the scalability test in two ways: finding how the model's runtime varies with population size and forecast horizon, and comparing the run time with a Mesa equivalence, which in our case is the unoptimised model. Results show that runtime grows linearly with respect to the number of simulated agents (\textbf{Figure 7}), aligning with the models mentioned above, and also with respect to the simulation length. The model is 200-250 times faster than the Mesa equivalence, reaching a simulation time of 1.40 seconds for 1000 simulated agents and 30 time steps, comparable to ~\cite{chopra2021deepabm} and ~\cite{chopradifferentiable}. 
\begin{figure}
    \centering
    \includegraphics[width=\linewidth]{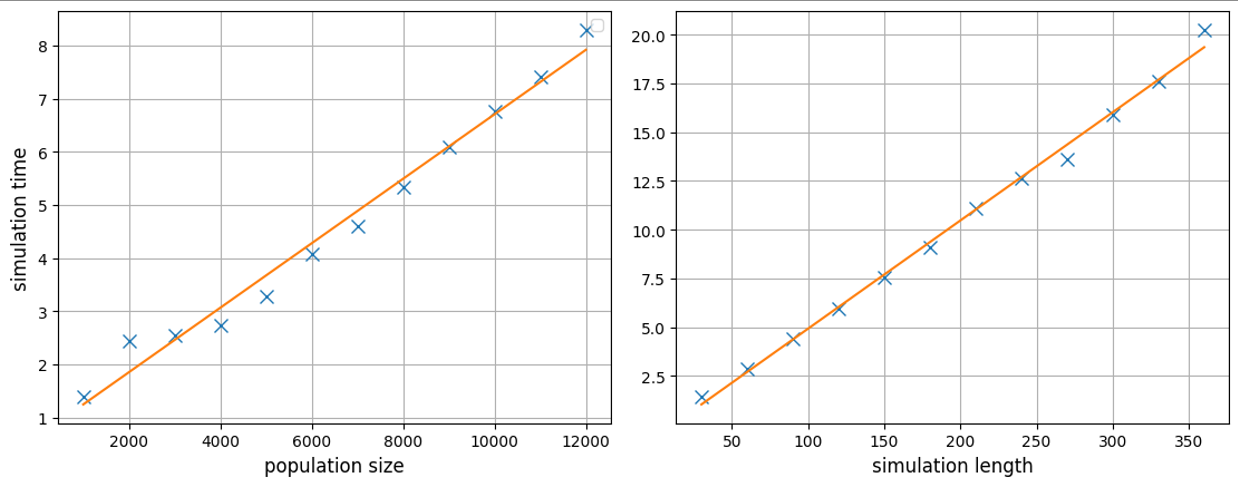}
    \caption{Runtime vs. number of simulated agents. A straight line of $y=0.3078x-0.4338$ can be fitted with a coefficient of determination $R=0.89$}
\end{figure}


\section{Conclusion}

We present DEpiABS, a scalable and interpretable structure-centric differentiable agent-based model (DABM). By enhancing feature granularity and rejecting the ANN component, DEpiABS fully leverages the potential of the DABM paradigm. It surpasses the state-of-the-art data-centric DABM in simulation fidelity and outperforms it with higher data efficiency while maintaining full mechanistic interpretability and comparable computational efficiency. We also propose and validate a z-score-based scaling method that decouples simulation cost from population size, enabling ABM forecasting for large-scale populations efficiently. As one of the first structure-centric DABMs, it demonstrates the competitiveness of this design philosophy over data-centric alternatives. 



\begin{acks}
    This research is supported by the Ministry of Education, Singapore, under its MOE AcRF Tier 2 Award MOE-T2EP20223-0003. Any opinions, findings and conclusions or recommendations expressed in this material are those of the author(s) and do not reflect the views of the Ministry of Education, Singapore.
\end{acks}



\clearpage
\bibliographystyle{ACM-Reference-Format} 
\bibliography{sample}


\clearpage
\appendix
\section{Model Details}


\subsection{Initialisation}
This model is initialised from 27 learnable parameters and 8 non-learnable parameters, illustrated in \textbf{Table 3} and \textbf{Table 4}. The learnable parameters regarding financial and supply states can also be manually configured from census data. Using these parameters, the model is initialised as configurations and states in \textbf{Tables 9-12}.
\begin{table*}[h]
    \centering
    \begin{tabular}{|p{2cm}|p{6cm}|}
        \hline
        \textbf{symbol} & \textbf{meaning} \\
        \hline
        $P$ & model population \\
        \hline
        $S_0, A_0, I_0, D_0$ & The initial numbers of agents under each health label \\
        \hline
        $T$ & The simulation length. \\
        \hline
        $n$ & The number of facility clusters. \\
        \hline
        $k$ & The steepness parameter of the logistic function. \\
        \hline
    \end{tabular}
    \caption{Non-learnable parameters}
\end{table*}
\begin{table*}[h]
    \centering
    \begin{tabular}{|p{2cm}|p{6cm}|}
    \hline
    \textbf{symbol} & \textbf{meaning}\\
    \hline
    $c$ & The price of one quota of supplies. \\
    \hline
    $sup_{\mu}$ & The mean initial supplies. \\
    \hline
    $sup_{\sigma}$ & The standard deviation of initial supplies. \\
    \hline  
    $sav_{\mu}$ & The mean disposable cash and savings. \\
    \hline
    $sav_{\sigma}$ & The standard deviation of disposable cash and savings. \\
    \hline    
    $bill_{\mu}$ & The mean monthly bill. \\
    \hline
    $bill_{\sigma}$ & The standard deviation of monthly bill. \\
    \hline   
    $\eta_\mu$ & The mean number of days of absence that triggers a salary cut. \\
    \hline
    $\eta_\sigma$ & The standard deviation of the number of days of absence that triggers a salary cut. \\
    \hline
    $prob$ & The mean probability of the most probable action under the highest emergency. \\
    \hline
    $dprob$ & The proportion of probability of the most probable action under the second-highest emergency to the most. \\
    \hline
    $ddprob$ & The proportion of probability of the most probable action under the third highest emergency to the second. \\
    \hline
    $prob_\sigma$ & The standard deviation of the probability of the most probable action under the highest emergency. \\
    \hline   
    $sal_{\mu}$ & The mean salary. \\
    \hline
    $sal_{\sigma}$ & The standard deviation of salary. \\
    \hline  
    $m$ & The probability of agents' encounters \\
    \hline
    $\beta$ & The probability of successful transmission. \\
    \hline
    $\theta$ & The age-impact proportion on symptom progression. \\
    \hline
    $age_\mu, age_\sigma$ & The mean and standard deviation of age distribution. \\
    \hline
    $\gamma$ & The probability of mutation. \\
    \hline
    $p$ & The probability of mutation-driven immunity escape. \\
    \hline
    $\sigma$ & The initial standard deviation of symptom progression. \\
    \hline
    $\mu$ & The initial mean of symptom progression. \\
    \hline
    $\tau$ & The mean judgement threshold between emergent and concerning epidemic severity as infected proportion. \\
    \hline
    $d\tau$ & The proportion of judgement threshold between concerning and under-control epidemic severity as infected proportion to the higher threshold. \\
    \hline
    $\zeta$ & The mean judgement threshold between emergent and concerning epidemic severity as the death rate. \\
    \hline
    $d\zeta$ & The proportion of judgement threshold between concerning and under-control epidemic severity as the death rate to the higher threshold. \\
    \hline
    $\delta$ & The mean judgement threshold for proportion of asymptomatic infection. \\
    \hline
    $d\delta$ & The proportion of judgement threshold between concerning and under-control epidemic severity as the asymptomatic proportion to the higher threshold. \\
    \hline
    $j_\sigma$ & The standard deviation of judgement thresholds. \\
    \hline      
    \end{tabular}
    \caption{Learnable parameters}
\end{table*}
\begin{table*}[h]
    \centering
    \begin{tabular}{|p{2cm}|p{6cm}|p{6cm}|}
        \hline
        \textbf{symbol} & \textbf{meaning} & \textbf{initialisation}\\
        \hline
        $[T]$ & The discrete time horizon & $T$ \\
        \hline
        $M$ & The map as facility set & $(m, n, c)$ \\
        \hline
    \end{tabular}
    \caption{Society configurations}
\end{table*}
\begin{table*}[h]
    \centering
    \begin{tabular}{|p{2cm}|p{6cm}|p{6cm}|}
        \hline
        \textbf{symbol} & \textbf{meaning} & \textbf{initialisation}\\
        \hline
        $\beta$ & The probability of transmission & $\beta$ \\
        \hline
        $\gamma$ & The probability of mutation & $\gamma$ \\
        \hline
        $p$ & The probability of mutation-driven immunity escape & $p$ \\
        \hline
        $\mu_t, \sigma_t$ & The mean and standard deviation of severity change & $\mu$, $\sigma$ \\
        \hline
        $mutate_t$ & The indicator of whether mutation happens at time step $t$. & 0 \\
        \hline
    \end{tabular}
    \caption{Epidemic configurations and states}
\end{table*}
\begin{table*}[h]
    \centering
    \begin{tabular}{|p{2cm}|p{6cm}|p{6cm}|}
        \hline
        \textbf{symbol} & \textbf{meaning} & \textbf{initialisation}\\
        \hline
        $j$ & The identity number of the agents & $j\in[1,P]$\\
        \hline
        $sal^{j}$ & salary & $sal^{j}\sim\mathcal{N}(sal_\mu, sal_\sigma)$\\
        \hline
        $dsal^{j}$ &  salary left after salary cut & $sal^{j}dsal^{j'}\sim\mathcal{U}(0, 1)$\\
        \hline
        $age^{j}$ & The age of the agents & $age^{j}\sim\mathcal{U}(18, 65)$. \\
        \hline
        $\theta^{j}$ & The age-impact factor & $\theta\cdot age^{j}$. \\ 
        \hline
        $bill^{j}$ & The monthly bill & $bill^{j}\sim\mathcal{N}(bill_\mu, bill_\sigma)$ \\
        \hline
        $\eta^{j}$ & The number of days of absence that triggers salary cut & $\eta^{j}\sim\mathcal{N}(\eta_\mu, \eta_\sigma)$\\
        \hline
        $\bar{\eta}^{j}$ & Thresholds of proportion of $\eta^{j}$ that drives absenteeism prevention. & $\eta^{j},d\eta^{j}\sim\mathcal{U}(0, 1)$ \\
        \hline
        $\bar{sup}_1^{j}, \bar{sup}_2^{j}$ & Thresholds between emergency levels 1 and 2, 2 and 3 for supply stats. & $\bar{sup}_1^{j}=0$, $\bar{sup}_2^{j}=sup_\mu$ \\
        \hline
        $\bar{s}_1^{j}, \bar{s}_2^{j}$ & Thresholds between emergency levels 1 and 2, 2 and 3 for symptom severity. & $\bar{s}_1^{j}=1$, $\bar{s}_2^{j}=4$ \\
        \hline
        $\tau_1^{j}$, $\tau_2^{j}$ & Thresholds between emergency levels 1 and 2, 2 and 3 for infected proportion. & $\tau_2^{j}\sim\mathcal{N}(\tau, j_\sigma)$, $\tau_1^{j}=\tau_2^{j}d\tau_2^{j}$ \\
        \hline
        $\zeta_1^{j}$, $\zeta_2^{j}$ & Thresholds between emergency levels 1 and 2, 2 and 3 for death rate. & $\zeta_2^{j}\sim\mathcal{N}(\zeta, j_\sigma)$, $\zeta_1^{j}=\zeta_2^{j}d\zeta_2^{j}$ \\
        \hline
        $\delta_1^{j}$, $\delta_2^{j}$ & Thresholds between emergency levels 1 and 2, 2 and 3 for asymptomatic proportion in infected population. & $\delta_2^{j}\sim\mathcal{N}(\delta, j_\sigma)$, $\delta_1^{j}=\delta_2^{j}d\delta_2^{j}$ \\
        \hline
        $prob_{ik}^{j}$ & The probabilities representing behavioural patterns under three different emergency levels. & \textbf{(16)} \\
        \hline
        $l^{j}$ & The work address assignment and market, hospital preferences. & randomly allocated from $M$ \\
        \hline
    \end{tabular}
    \caption{Agent configurations}
\end{table*}
\begin{table*}[h]
    \centering
    \begin{tabular}{|p{2cm}|p{6cm}|p{6cm}|}
        \hline
        \textbf{symbol} & \textbf{meaning} & \textbf{initialisation}\\
        \hline
        $d_t^{j}$ & The decision made by the agent. & N/A \\
        \hline
        $l_t^{j}$ & The location chosen by the agent. & N/A \\
        \hline
        $sup_t^{j}$ & Available supplies in daily quotas. & $sup_t^{j}\sim\mathcal{N}(sup_\mu, sup_\sigma)$ \\
        \hline
        $sav_t^{j}$ & Disposable cash and savings. & $sav_t^{j}\sim\mathcal{N}(sav_\mu, sav_\sigma)$ \\
        \hline
        $absent_t^{j}$ & The number of days of absence from work of the current month. & 0 \\
        \hline
        $t_{inc}^{j}, t_{sym}^{j}$ & The lengths of the incubation and symptomatic periods of the current infection the agent is undergoing at time step $t$. & \textbf{(1)} \\
        \hline
        $s_t^{j}$ & The symptom severity of the agent at time step $t$. & 0 \\
        \hline
        $t_{0_t}^{j}$ & The time step at which the current infection of the agent started. & 0 \\
        \hline
        $immune_{t}^{j}$ & The immunity level of the agent. & 1 \\
        \hline
        $h_t^{j}$ & The health class of the agent. & $j, S_0, A_0, I_0, D_0$ \\
        \hline
    \end{tabular}
    \caption{Agent states}
\end{table*}

\subsection{Society}
The supply state of agent $j$ is updated by
\begin{equation}
    \begin{split}
        &sup_t^{j}=sup_{t-1}^{j}+{d_t^{j}}_3dsup_t^{j}
        \\&
        dsup_t^{j}\sim\mathcal{U}(sup_\mu, sup_\sigma)
    \end{split}
\end{equation}
and the financial states are updated by
\begin{equation}
    \begin{split}
        \\sav_t^{j}&=sav_{t-1}^{j}-d_t^{j}dsup_t^{j}\cdot c
        \\&
        +\rho(t;0, 30)\cdot(sal_t^{j}(dsal_t^{j}\cdot\mathbf{I}(absent_t^{j}>\eta^{j})
        \\&
        +\mathbf{I}(\eta^{j}>absent_t^{j}))-bill^{j})
        \\absent_t^{j}&=(absent_{t-1}^{j}+(1-(\rho(t; 6, 7) + \rho(t; 0, 7)))
        \\&\cdot(1-{d_t^{j}}_2))(1-\rho(t; 0, 30))
    \end{split}
\end{equation}

\subsection{Epidemic}
Health classes are updated via
\begin{equation}
    \begin{split}
        &{h_t}_1={h_{t-1}}_1-infect_t+recovery_t
        \\&
        {h_t}_2=infect_t+\mathbf{I}(\mathbf{1}>((s_t^{j})_j)\circ({h_{t-1}}_2+{h_{t-1}}_3))
        \\&
        {h_t}_3=\mathbf{I}(((s_t^{j})_j-\mathbf{1})\circ(6\cdot\mathbf{1}-(s_t^{j})_j)>\mathbf{0})
        \\&
        {h_t}_4=\mathbf{I}((s_t^{j})_j)>6\cdot\mathbf{1})
    \end{split}
\end{equation}
where $recovery_t$ represents agents recovering at this time step, computed by
\begin{equation}
    \begin{split}
        &per_t=\mathbf{I}(-\mathbf{1}>(s_t^{j})_j)\mathbf{I}(\mathbf{1}>per_{t-1})\cdot({h_{t-1}}_2+{h_{t-1}}_3)
        \\&
        er_t=\mathbf{I}((-\mathbf{1}>(s_t^{j})_j))\cdot({h_{t-1}}_2+{h_{t-1}}_3)\cdot per_{t-1}
        \\&
        r_t=\mathbf{I}({t_I}_t-({t_i}_t+{t_s}_t))\cdot(({h_{t-1}}_2+{h_{t-1}}_3)-per_{t-1})
        \\&
        recovery_t=er_t+r_t
    \end{split}
\end{equation}
Here, $per_t, er_t,$ and $r_t$ are mediating variables representing agents potentially early recovering, i.e. had non-positive symptom severity this time step but not the time step before that, agents early recovering, and agents recovering due to the elapse of incubation and symptomatic periods. Empirical studies indicate a three-day maximum of asymptomatic duration for patients to recover; we therefore chose a smaller average value.

The symptom dynamics follow a random walk with increments drawn from a Normal distribution, whose mean and standard deviation vary when a mutation occurs, and are age-sensitive. With symbols in \textbf{Table 5}, it can be modelled as 
\begin{equation}
    \begin{split}
        &s_t^{j} = \sum_{t'=t_{0_t}^{j}}^{t} \frac{\theta\cdot age^{j} \cdot ds_{t'}^{j}}{immune_{t'}^{j}}
        \\&
        ds_{t'}^{j}\begin{cases}
            \sim \mathcal{N}(\mu_{t'}, \sigma_{t'})\text{ if }t'-t_{0_t}^{j}>t_{inc}^{j}\\
            =0\qquad\qquad\text{ otherwise}\\
        \end{cases}
    \end{split}
\end{equation} 
\begin{table*}[h]
    \centering
    \begin{tabular}{|p{2cm}|p{6cm}|}
        \hline
        symbol & meaning \\
        \hline
        $\theta$ & The factor of age sensitivity of symptom progression. \\
        \hline
        $\mu_{t'}, \sigma_{t'}$ & The mean and standard deviation of severity change at time step $t'$ \\
        \hline
        $age^{j}$ & The age of agent $j$. \\
        \hline
        $t_{inc}^{j}, t_{sym}^{j}$ & The lengths of the incubation and symptomatic periods of the current infection that agent $j$ is undergoing at time step $t$. \\
        \hline
        $s_t^{j}$ & The symptom severity of agent $j$ at time step $t$ \\
        \hline
        $t_{0_t}^{j}$ & The time step at which the current infection of the agent $j$ started. \\
        \hline
        $ds_{t'}^{j}$ & The change in symptom severity of agent $j$ at time step $t'$. \\
        \hline
        $immune_{t'}^{j}$ & The immunity level of agent $j$ at time step $t$. \\
        \hline
    \end{tabular}
    \caption{Symbols involved in symptom progression}
\end{table*}

Mutation and mutation-driven immune escape are modelled by
\begin{equation}
    \begin{split}
        &mutate_t\sim \text{Ber}(\gamma)
        \\&
        immune_t^{j}=immune_{t-1}^{j}-mutate_t\cdot dimmune_t^{j}
        \\&
        dimmune_t^{j}\sim\text{Ber}(p)
    \end{split}
\end{equation}

\subsection{Population}
\textbf{Table 6} illustrates all the mediating variables involved in this component. The decision probabilities initialised by
\begin{equation}
    \begin{split}
        &prob_1=prob
        \\&
        prob_2=prob\cdot dprob
        \\&
        prob_3=prob\cdot dprob\cdot ddprob
        \\&
        prob_{i1}=prob_i
        \\&
        prob_{i2}=\frac{1}{2}(1-prob_{i1})=\frac{1}{2}-\frac{1}{2}prob_{i1}
        \\&
        prob_{i3}=\frac{3}{5}(1-prob_{i1}-prob_{i2})=\frac{3}{10}-\frac{3}{10}prob_{i1}
        \\&
        prob_{i4}=\frac{1}{5}-\frac{1}{5}prob_{i1}
    \end{split}
\end{equation}

The computation of the drive to stay at home is complex, as epidemic severity depends collectively on three aspects. Its design follows three rules:
\begin{itemize}
    \item \textbf{Even distribution:} Each drive level corresponds to an equal number of collective situations.
    \item \textbf{Higher-sum-higher-total:} The total drive increases with the sum of drives from all aspects.
    \item \textbf{Ranked influence:} The influences of the three aspects decrease in order: infected proportion, death rate, and asymptomatic proportion.
\end{itemize}
Even distribution ensures consistency with other informational components. Higher-sum-higher-total provides an intuitively valid configuration. Ranked influence prioritises infection prevalence, reflecting the epidemic’s spread, over the death rate, consistent with how COVID-19, though less fatal, caused global disruption, whereas Ebola, despite its high fatality, remained regionally contained. The asymptomatic proportion is treated as secondary. Drives from each aspect are represented as three-entry vectors, and the total drive from epidemic severity is determined as shown in \textbf{Table 7}.
\begin{table*}[h]
    \centering
    \begin{tabular}{|p{2cm}|p{6cm}|}
        \hline
        Total & Drives from aspects \\
        \hline
        1 & (1, 1, 1), (1, 1, 2), (1, 2, 1), (2, 1, 1), (1, 2, 2), (2, 1, 2), (2, 2, 1), (1, 1, 3), (1, 3, 1) \\
        \hline
        2 & (3, 1, 1), (1, 2, 3), (1, 3, 2), (2, 1, 3), (2, 2, 2), (2, 3, 1), (3, 1, 2), (3, 2, 1), (1, 3, 3) \\
        \hline
        3 & (2, 2, 3), (2, 3, 2), (3, 1, 3), (3, 2, 2), (3, 3, 1), (2, 3, 3), (3, 2, 3), (3, 3, 2), (3, 3, 3) \\
        \hline
    \end{tabular}
    \caption{Drives from epidemic severity and their correspondences to drives from each perspective}
\end{table*}

In this setup, level-1 drives correspond to perspective-drive sums of 3, 4, and 5 (when the drive from the infected proportion is below 3); level-2 drives correspond to sums of 5 (when the infected-proportion drive equals 3), 6, and 7 (when it equals 1); and level-3 drives correspond to sums of 7 (when the infected-proportion drive exceeds 1), 8, and 9. These assignments satisfy the principles of equal distribution and higher-sum-higher-total. For ranked influence results, we present \textbf{Table 8}, which indicates that the distribution of situations driven by the infected proportion is more skewed than those by death rate or asymptomatic proportion, implying a stronger influence. The implementation enumerates all possible combinations, and the overall mechanism for determining emergency levels and action plausibilities is summarised in \textbf{Algorithm 1}. With this, the behavioural rule is represented as \textbf{Algorithm 2}. In implementation, we hard-coded these rules by explicitly computing each different scenario, merging and eliminating scenarios that induce identical emergencies (e.g., (3, 3, 1), (3, 3, 2) and (3, 3, 3) for epidemic severity), and tensorising them.  
\begin{algorithm}
    \caption{Behavioural Rule}
    \begin{algorithmic}[1]
        \STATE \textbf{Input:} $\mathcal{T}_{t-1}, \mathcal{Z}_{t-1}, \Delta_{t-1}, s_{t-1}^{j}, absent_t^{j}, sav_t^{j}, sup_{t-1}^{j}$
        \STATE \textbf{Output:} $l_t^{j}$
        \STATE \textbf{Thresholds:} $\tau_1^{j}, \tau_2^{j}, \zeta_1^{j}, \zeta_2^{j}, \delta_1^{j}, \delta_2^{j}, \bar{s}_1^{j}=1, \bar{s}_2^{j}=4,$ \STATE $\hspace{5.5em} bill^{j}, \eta^{j}, \bar{\eta}^{j}, \bar{sup}_1^{j}=0, \bar{sup}_2^{j}=3$
        \STATE $\tilde{\tau}_{t_1}\leftarrow\mathcal{T}_{t-1}-\tau_1^{j}, \dots, \tilde{sup}_{t_1}^{j}\leftarrow sup_{t-1}^{j}-\bar{sup}_1^{j}$ 
        \STATE $\mathcal{A}_t^{j}, \alpha_t^{j}\leftarrow\textbf{Algorithm 1}$
        \STATE $i\leftarrow \mathcal{A}_t^{j}, (k_1, \dots, k_4)\leftarrow \alpha_t^{j}$ 
        \STATE $probs_t^{j}\leftarrow(prob_{i1}^{j}, \dots, prob_{i4}^{j})$
        \FOR{$l\in[1, 2, 3, 4]$}
            \STATE $prob_{t_l}^{j}=probs_{t_{k_l}}^{j}$
        \ENDFOR
        \STATE $d_t^{j}\sim\text{Categorical}((prob_{t_l}^{j})_{l\in[1, 4]})$
        \STATE $l_t^{j}\leftarrow L^{j}\cdot_{3}d_t^{j}$
        \RETURN $d_t^{j}$
    \end{algorithmic}
\end{algorithm}
\begin{algorithm}
    \caption{Emergency Level and Plausibilities}
    \begin{algorithmic}[1]
        \STATE \textbf{Input:} $ \tilde{\tau}_{t_1}^{j}, \tilde{\tau}_{t_2}^{j}, \tilde{\zeta}_{t_1}^{j}, \tilde{\zeta}_{t_2}^{j}, \tilde{\delta}_{t_1}^{j}, \tilde{\delta}_{t_2}^{j}, \tilde{s}_{t_1}^{j}, \tilde{s}_{t_2}^{j},$
        \STATE $\hspace{5.5em} \tilde{bill}_t^{j}, \tilde{\eta}_t^{j}, \tilde{sup}_{t_1}^{j}, \tilde{sup}_{t_2}^{j}, t$
        \STATE \textbf{Output:} $\mathcal{A}_t^{j}, \alpha_t^{j}$
        \FOR{$a\in[\tau, \zeta, \delta, s, sup]$}
            \IF{$\tilde{a}_{t_1}^{j}<0$}
                \STATE $\mathcal{A}_{t_a}^{j}\leftarrow1$
            \ELSIF{$\tilde{a}_{t_1}^{j}\geq0$ and $a_{t_2}<0$}
                \STATE $\mathcal{A}_{t_a}^{j}\leftarrow2$
            \ELSE
                \STATE $\mathcal{A}_{t_a}^{j}\leftarrow3$
            \ENDIF
        \ENDFOR
        \IF{$\rho(t; 6, 7)\cdot\rho(t, 0, 7)=0$}
            \STATE $\mathcal{A}_{t_{finance}}^{j}\leftarrow1$
        \ELSE
            \IF{$\tilde{bill}_t^{j}<0$ \textbf{AND} $\tilde{\eta}_t^{j}<0$}
                \STATE $\mathcal{A}_{t_{finance}}^{j}\leftarrow1$
            \ELSIF{$\tilde{bill}_t^{j}\geq0$ \textbf{OR} $\tilde{\eta}_t^{j}\geq0$}
                \STATE $\mathcal{A}_{t_{finance}}^{j}\leftarrow2$
            \ELSIF{$\tilde{bill}_t^{j}\geq0$ \textbf{AND} $\tilde{\eta}_t^{j}\geq0$}
                \STATE $\mathcal{A}_{t_{finance}}^{j}\leftarrow3$
            \ENDIF
        \ENDIF 
        \STATE $\mathcal{A}_{t_{epidemic}}^{j}\leftarrow\textbf{Table 10}$
        \STATE $\mathcal{A}_{t_{health}}^{j}\leftarrow\mathcal{A}_{t_s}^{j}$
        \STATE $\mathcal{A}_{t_{supply}}^{j}\leftarrow\mathcal{A}_{t_{sup}}^{j}$
        \STATE $\mathcal{A}_{t_1}^{j}\leftarrow\text{min}(3, \mathcal{A}_{t_{epidemic}}^{j}+\text{min}(1, \mathcal{A}_{t_{health}}^{j}))$
        \STATE $\mathcal{A}_{t_2}^{j}\leftarrow\mathcal{A}_{t_{finance}}^{j}$
        \STATE $\mathcal{A}_{t_3}^{j}\leftarrow\mathcal{A}_{t_{supply}}^{j}$
        \STATE $\mathcal{A}_{t_4}^{j}\leftarrow\mathcal{A}_{t_{health}}^{j}$
        \STATE \textbf{sort} $\alpha_t=[1, 2, 3, 4]$ with respect to $[\mathcal{A}_{t_1}^{j}, \mathcal{A}_{t_2}^{j}, \mathcal{A}_{t_3}^{j}, \mathcal{A}_{t_4}^{j}]$
        \STATE $\mathcal{A}_t^{j}\leftarrow{\mathcal{A}_{t_{\alpha_t}}}_0^{j}$
        \RETURN $\mathcal{A}_t^{j}, \alpha_t^{j}$
    \end{algorithmic}
\end{algorithm}
\begin{table*}[h]
    \centering
    \begin{tabular}{c|ccc|ccc|ccc}
        \toprule
         & \multicolumn{3}{c|}{$\tau$} & \multicolumn{3}{c|}{$\zeta$} & \multicolumn{3}{c}{$\delta$} \\
        \cmidrule(lr){2-4} \cmidrule(lr){5-7} \cmidrule(lr){8-10}
         & 1 & 2 & 3 & 1 & 2 & 3 & 1 & 2 & 3 \\
        \midrule
        1 & 6 & 3 & 0 & 5 & 3 & 1 & 5 & 3 & 1 \\
        2 & 3 & 3 & 3 & 3 & 3 & 3 & 3 & 3 & 3 \\
        3 & 0 & 3 & 6 & 1 & 3 & 5 & 1 & 3 & 5 \\
        \bottomrule
    \end{tabular}
    \caption{The number of situations that give each level of drive from each perspective for the three emergency levels}
\end{table*}
\begin{table*}[h]
    \centering
    \begin{tabular}{|p{2cm}|p{6cm}|}
        \hline
        symbol & meaning \\
        \hline
        $\tilde{sup}_{t_1}^{j}, \dots, \tilde{\eta}_{t_2}^{j}$ & Indicators of emergency level in each aspect of information. \\
        \hline
        $\mathcal{A}_{t_a}^{j}$ & The drive from perspective $a$. \\
        \hline
        $\mathcal{A}_{t_i}^{j}$ & The drive to action $i$. \\
        \hline
        $\mathcal{A}_t^{j}$ & The emergency level of the current situation. \\
        \hline
        $\alpha_t^{j}$ & The indices of plausibility from high to low of each action. \\
        \hline
        $probs_t^{j}$ & The probability set used given the emergency level. \\
        \hline
        $prob_{t_l}^{j}$ & The probability of choosing $lth$ action. \\
        \hline
        $probs_{t_{k_l}}$ & The probability mass assigned to the $lth$ action given its plausibility. \\
        \hline
    \end{tabular}
    \caption{Symbols involved in the Population component}
\end{table*}


\section{Experiment Details}

\subsection{Implementation}
The experiments were run using Python 3.11.0. The main body of the model and the calibration and forecasting experiments are implemented using Pytorch 2.5.0 and CUDA 1.2.4. The traditional version of the model is implemented using Mesa~\cite{python-mesa-2020}. All experiments are executed on the CPU AMD Ryzen 7 7435HS with 20 GB of RAM, and the GPU NVIDIA GeForce RTX 4060. 

\subsection{Calibration}
We followed the same processing, calibration and forecasting procedure as the baseline~\cite{chopradifferentiable}. Simulations for the remaining nine counties are shown in \textbf{Figures 8 \& 9} for COVID-19 and influenza, respectively. To mitigate the numerous saddle points and local minima induced by the complex model dynamics, we initialised gradient descent with a large learning rate and manually annealed it from 0.1 to 0.001.

\begin{figure*}[h]
    \centering
\includegraphics[width=\linewidth]{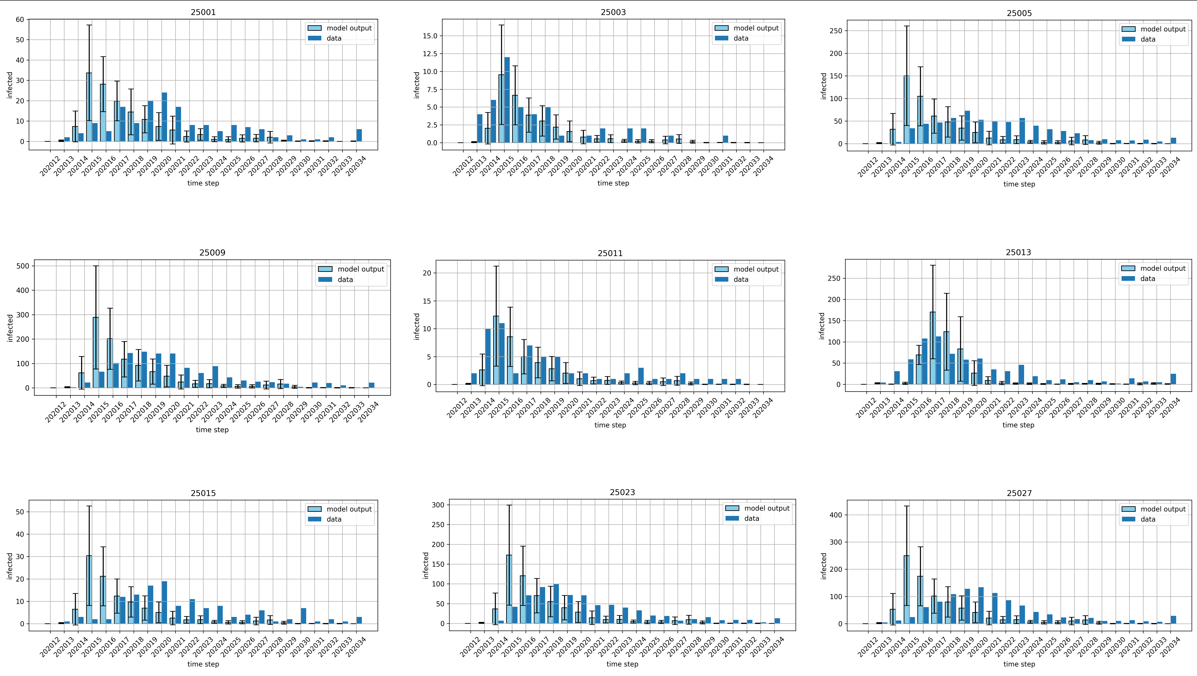}
    \caption{Forecasted and real COVID mortalities in the other 9 counties}
\end{figure*}
\begin{figure*}[h]
    \centering
    \includegraphics[width=\linewidth]{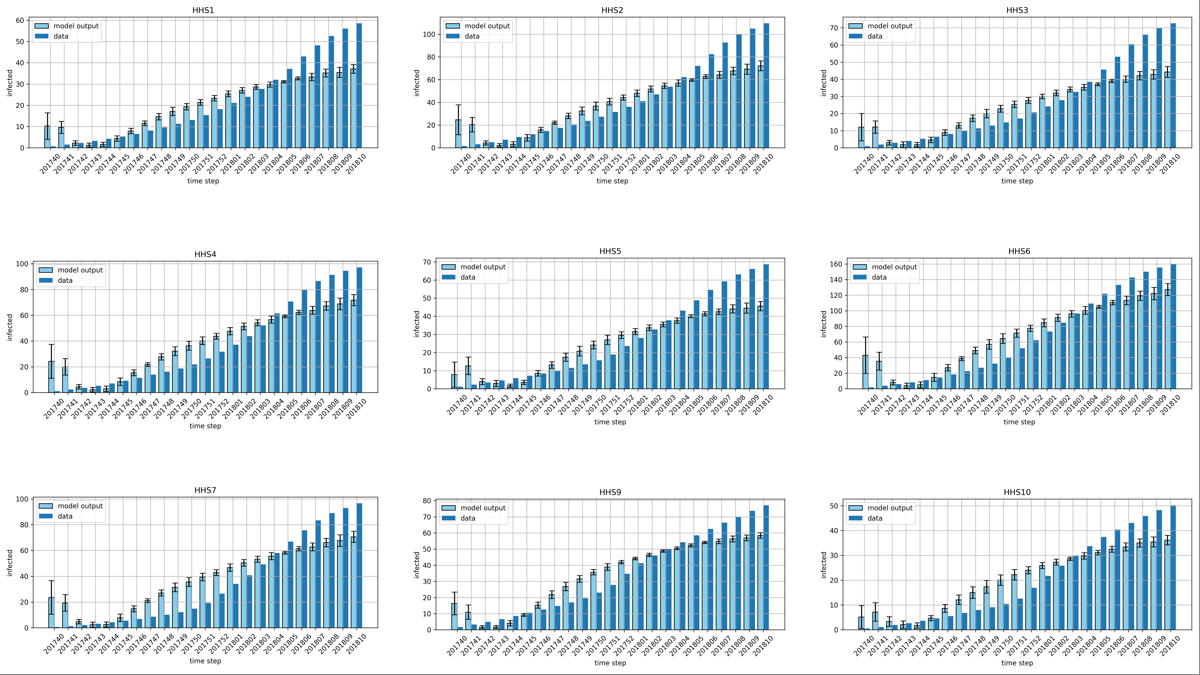}
    \caption{Forecasted and real ILIs in the other 9 counties}
\end{figure*}

\end{document}